\documentclass[preprint2]{aastex}

\catcode`\@=11
\def\gapprox{\mathrel{\mathpalette\@versim>}}
\def\lapprox{\mathrel{\mathpalette\@versim<}}
\def\@versim#1#2{\lower2.9truept\vbox{\baselineskip0pt\lineskip0.5truept
    \ialign{$\m@th#1\hfil##\hfil$\crcr#2\crcr\sim\crcr}}}
\def\sun{\ifmmode\odot\else$\odot$\fi}

\newcommand{\vdag}{(v)^\dagger}

\shorttitle{Fast Radio Bursts: Constraints on the Dispersing Medium}
\shortauthors{Dennison}

\begin{document}


\title{Fast Radio Bursts: Constraints on the Dispersing Medium}


\author{Brian Dennison}
\affil{Department of Physics, University of North Carolina-Asheville,
    Asheville, NC 28804}
\email{dennison@unca.edu}




\begin{abstract}
Fast radio bursts appear to exhibit large dispersion measures, typically exceeding any expected galactic interstellar contribution, especially along the moderate to high-galactic-latitude directions in which such events have been most often observed. The dispersions have been therefore interpreted as extragalactic, leading to the inference that the sources of the bursts are at Gpc distances. This then implies that the bursts are extremely energetic events, originating from quite small volumes (due to the millisecond burst durations). To circumvent the energetic difficulties, \citet{l1} propose that the bursts are produced by flares near the surfaces of M stars or contact binaries within a local volume of the galaxy. Most of the dispersion would then occur in the overlying stellar coronae. With the dispersion concentrated in a relatively high density region, the quadratic dispersion approximation breaks down as the plasma frequency is comparable to (although less than) the propagation frequency. The observed dispersion curves are closely quadratic, however, consistent with a low density medium, ruling out this model. Because any model invoking local galactic sources would require a concentrated high density dispersing medium, it appears highly likely that the dispersions occur in the intergalactic medium. This medium, probably containing most of the baryon content of the universe, is expected to be highly structured on large scales. Hot gas within clusters and especially groups of galaxies may contribute significantly to the observed dispersion. Optical and X-ray observations, including redshifts and combined with cosmological dispersion measures, can probe the distribution of the IGM and determine how much of it lies outside the X-ray luminous concentrations seen in clusters and groups.
\end{abstract}





\section{Introduction}

Fast radio bursts (FRBs) are apparently solitary events of millisecond monochromatic duration, detected at GHz frequencies. Because most events exhibit quadratic frequency-dependent delay, the apparent dispersion has been ascribed to propagation in cold plasma, leading to the tentative conclusion that they are extraterrestrial in origin \citep{l2,t1,k3}. 
Some events, dubbed perytons, are suspected, however, of possibly having a terrestrial origin \citep{b4,s1}. In what follows I regard the six events reported by \citet{l2} (FRB 010824), \citet{k3} (FRB 010621), and \citet{t1} (FRBs 110220, 110627, 110703, and 120127) as extraterrestrial. These events show precise quadratic drifts in delay. The inferred dispersion measures (DMs) are quite large, ranging from 375 cm$^{-3}$ pc (FRB 010824) to 1072 cm$^{-3}$ pc (FRB 110703). These six events do not show any clear correlation with galactic latitude, although FRB 010621 was observed at low galactic latitude ($b = -4^\circ$). The other five FRBs were seen at latitudes larger than $40^\circ$. Although \citet{b2} argue that the dispersion observed in FRB 010621 originates in the interstellar medium of the Milky Way, such a conclusion is untenable for the other five FRBs observed at high latitudes.
 
The large dispersions of the high latitude FRBs have thus been ascribed to an ionized intergalactic medium \citep{l2,t1}. This places the sources of the bursts at moderate cosmological distances ($\lapprox 4$ Gpc). 
Two bursts (FRBs 010824 and 110220) also appear to exhibit temporal broadening caused by scattering by density inhomogeneities along the lines of sight \citep{l2,t1}. The observed frequency of bursts, combined with the very limited sky coverage that led to these detections, indicates that $\approx 10^4$ bursts with fluence $\approx 3$ Jy ms reach the Earth daily \citep{t1}.

An extragalactic origin for the bursts necessarily requires quite large radiated energy, even if relativistic beaming is invoked. Incoherent synchrotron emission in particular is problematic because of the very large brightness temperatures that are inferred. Coherent emission involving bunched electrons has been considered by \citet{k2}. 

A variety of physical mechanisms have been considered as extragalactic burst sources, including black hole evaporation \citep{k3}, soft gamma ray repeaters \citep{t1,p1,k4}, collapsing supramassive neutron stars \citep{f1,z1,k4}, 
binary white dwarf mergers \citep{k1}, synchrotron maser emission from relativistic shocks \citep{l3}, and neutron star mergers \citep{t2}. 

\section{Dispersion in Stellar Coronae} 

To alleviate the difficulties inherent in high energy sources, \citet{l1} propose that the bursts originate with a local stellar population within the galaxy. The brightest events, i.e those that associated with stars near the Sun, would then be expected to exhibit a broad latitude distribution consistent with that observed. \citet{l1} propose that the bursts are caused by flares near the surfaces of M stars or contact binaries. Most of the dispersion, and scattering if present, would then occur in the overlying stellar corona, whether through the steady state coronal structure or ionized material ejected in the flare event. An essential feature of such local models is that they concentrate the ionized gas responsible for the observed dispersion in the immediate vicinity of the source itself.

To produce the observed dispersion the ionized gas is necessarily moderately dense. As \citet{l1} point out, the typical density of this gas would be $\approx 10^{10}\ {\rm cm}^{-3}$, with a plasma frequency $\nu_p\approx 0.9\ {\rm GHz}$. Although below the frequency at which events have been observed, this will result in an arrival time  frequency-dependence measurably different from that observed.

The group velocity of electromagnetic waves in a cold plasma is given by
\begin{equation} 
V_g = c\sqrt{1 - \big({\omega_p\over\omega}\big)^2}
\end{equation} 

Expanding this expression yields
\begin{equation} 
V_g = c\Big[ 1 - {1\over 2} \big({\omega_p\over\omega}\big)^2 - {1\over 8} \big({\omega_p\over\omega}\big)^4 - {1\over{16}} \big({\omega_p\over\omega}\big)^6 + ...\Big]
\end{equation} 

Conventional treatments of pulsar dispersion utilize just the first two terms in the expansion because all higher order terms are negligible at the plasma frequency of the interstellar medium. Indeed the DMs obtained for the six known FRBs were evidently calculated  in this manner. The published dispersion curves do in fact display a close correspondence to the quadratic approximation. 

Models invoking concentrated dispersion and therefore high densities of ionized gas necessitate the consideration of what might be termed {\sl strong dispersion}, in which the dispersion curve displays marked deviation from the quadratic delay indicative of low gas densities. In such cases either higher order terms, or better yet the exact form of the of the group velocity expression (1) must be used.

Consider a flare emitted  near the surface of  a star. The electron density in the overlying corona or wind producing the observed dispersion could be modeled as
\begin{equation} 
n = n_0 \Big({R_0\over r}\Big)^2,
\end{equation} 
where $R_0$ is the stellar radius. In this case, the dispersive pulse delay can be worked out exactly for a radial trajectory; it is given by
\begin{equation} 
\Delta t(\omega ) = {R_0\over c}\Big( 1 - \sqrt{1 - \big({\omega_{p0}\over\omega}\big)^2}\Big),
\end{equation} 
where $\omega_{p0}$ is the plasma frequency at the bottom of the corona, i.e. where $n = n_0$.

Consider the first event reported by \citet{l2} for which ${\rm DM} = 375$ cm$^{-3}$ pc, of which only $\approx 25$ cm$^{-3}$ pc is attributable to galactic interstellar gas. If the remaining DM of $\approx 350$ cm$^{-3}$ pc is produced within the stellar corona of a star with $R_0\approx R_\sun$, then $n_0 \approx 1.6\times 10^{10}$ cm$^{-3}$ and $\nu_{p0} = \omega_{p0}/(2\pi) \approx 1.1$ GHz. In this case the dispersion must be regarded as {\sl strong}. The published dynamic spectrum, however, closely fits the quadratic, i.e. weak dispersion, approximation. 
Using the above parameters, one finds that the shape of the predicted pulse delay curve differs markedly from that observed, resulting in a discrepancy in delay of 200 ms across the band (1.25 GHz -- 1.50 GHz). 
More generally, attempting to fit the observed dynamic pulse spectrum to equation (4) 
would yield $\omega_{p0} << 2\pi\times 1$ GHz and $R_0 >> R_\sun$.
This conclusion is not significantly altered if the compact dispersing medium has a different density distribution.  If this medium is  concentrated on scales of stellar radii {\sl strong dispersion} applies. The somewhat smaller radius of an M-star only makes the problem worse. An additional complication occurs because significant refraction in the corona will modify ray trajectories. 

Two of the events described by \citet{t1} are reported with fitted dispersion indices, $\alpha$, and associated errors, such that the arrival delay varies with frequency as $\nu^{-\alpha}$. This conveniently allows setting limits to the density of the dispersing gas. In general, the arrival time is given by
\begin{equation} 
t = {1\over c}\int {{dx}\over{\sqrt{1 - ({\omega_p\over\omega})^2}}},
\end{equation} 
where the integral is along the line of sight. Expanding to fourth order in frequency and solving for the frequency-dependent delay gives
\begin{equation} 
t - t_0 = {1\over 2}\int \big({\omega_p\over\omega}\big)^2 dx + {3\over 8}\int \big({\omega_p\over\omega}\big)^4 dx,
\end{equation} 
where $t_0$ is the total path length divided by the speed of light. The two  terms on the right are proportional to the electron column density $N_1 = \int n dx$ and the column-integrated density-squared $N_2 = \int n^2 dx$. The electron column density is of course usually expressed as the dispersion measure. If the gas is sufficiently dense the second integral will modify the otherwise expected $\nu^{-2}$ dependence of the pulse arrival time. 
The observed pulse dynamic spectra were fitted to the effective profile $t - t_0 = K \omega^{-\alpha}$ by \citet{t1}.  Consider a pulse observed over a relatively narrow band centered at $\omega_0 = 2\pi (1.35\times 10^9\ {\rm Hz})$. We then require that the fitting function and the actual profile (equation 6) yield identical values for $(t-t_0)$ and $d(t-t_0)/d\omega$ at $\omega = \omega_0$. This results in a relation between the fitted value of $\alpha$ and the density integrals:
\begin{equation} 
{{N_2}\over{N_1}} = {{\alpha - 2}\over{4-\alpha}}\ {{m\omega_0^2}\over{4\pi e^2}}
\end{equation} 

Table 1 gives results for FRBs 110220 and 110703. Both pulses are consistent with $\alpha = 2.00$ and therefore $N_2 = 0$. Here I use the 3-$\sigma$ limits to $\alpha$ to establish approximate upper bounds on $N_2$. In both cases the values for $N_1$ are obtained from the DMs minus the expected small galactic contribution. Two possibilities for the dispersing medium are considered: (i) It is uniform, in which case the electron density is given by $n = N_2/N_1$; and (ii) it consists of a stellar corona/wind with radial density variation given by equation (3). In case (i) an upper limit to the electron density is obtained and a lower limit to the path length, $X$, through the uniform medium results (as constrained by the observed DM). In case (ii)  $N_1 = n_0R_0$ and $N_2 = {1\over 3} n_0^2 R_0$. This results in upper limits to the electron density at the base the corona and lower limits the stellar radii.

\begin{table*}
 \centering
 \begin{minipage}{140mm}
  \caption{Strong Dispersion Constraints on two Fast Radio Bursts.}
   \begin{tabular}{@{}llll@{}}
  \hline 
  FRB & 110220 & 110703 \\
  ${\rm DM}_{EG}$ & 910 & 1072 \\
  $N_1$ & $2.8\times 10^{21}$ cm$^{-2}$ & $3.3\times 10^{21}$ cm$^{-2}$ \\
  $\alpha$ & $2.003\pm 0.006$ & $2.000\pm 0.006$ \\
  $\alpha_{3\sigma}$ & $< 2.021$ & $< 2.018$ \\
  $N_2$ & $< 6.74\times 10^{29}$ cm$^{-5}$ & $< 6.80\times 10^{29}$ cm$^{-5}$ \\
  $n_e$ & $< 2.4\times 10^8$ cm$^{-3}$ & $< 2.1\times 10^8$ cm$^{-3}$ & Case i (Uniiform)\\
  $X$ & $> 1.2\times 10^{13}$ cm & $> 1.6\times 10^{13}$ cm& Case i (Uniform)\\
  $n_0$ & $< 7.2\times 10^8$ cm$^{-3}$ & $< 6.2\times 10^8$ cm$^{-3}$ & Case ii (Corona)\\
  $R_0$ & $> 3.9\times 10^{12}$ cm & $> 5.4\times 10^{12}$ cm & Case ii (Corona)\\

\hline
\end{tabular}
\end{minipage}
\end{table*}

Not surprisingly, the stellar radii are larger than those of main sequence stars, thus ruling out the scenarios considered by \citet{l1}. Conceivably such pulses could originate near the surfaces of post-main-sequence giant stars. Such stars, however, often have large mass loss rates resulting in winds that could produce DMs several orders of magnitude in excess of those observed, depending on the ionization state of the outflowing gas. Giant stars being highly luminous could be easily identified in an optical search of the fields of known FRBs.

\section{Dispersion in the Intergalactic Medium}

The above considerations suggest that most of the observed dispersion originates outside the Milky Way Galaxy. One is thus forced to tentatively concur with the original interpretations that these events are extragalactic and quite possibly cosmologically distant. These existing interpretations tend to thus place the dispersion in the general intergalactic medium (IGM). If correct this has major implications because (i) the events themselves must be highly energetic, even if relativistic beaming is involved; and (ii) such observations could provide a unique and extremely valuable probe of the IGM.

\citet{t1} have examined the possibility that some significant fraction of the observed dispersions are produced in the host galaxies. As these authors have pointed out, however, 
this seems unlikely as every case would involve either
propagation through a galactic center or the disk of a spiral galaxy observed nearly edge-on, selection effects notwithstanding.

The existing distance estimates based on dispersion measures \citep{l2,t1} assume a uniform IGM having essentially the baryon content of the universe. It is likely however that much of the IGM is concentrated within large scale structure as outlined by groups and clusters of galaxies, and on larger scales, by filaments. The denser concentrations of intergalactic gas are well observed as the X-ray emitting intracluster medium and intragroup medium. \citet{j1} summarize the physical parameters of X-ray emitting gas in clusters. The radial column density of ionized gas integrated out to eight core radii and averaged over 521 observed clusters gives  ${\rm DM} \approx 800$ cm$^{-3}$ pc. This is likely to be comparable to the DM expected for a line of sight passing through a cluster. The odds of encountering a rich cluster over a Gpc line of sight are small, however, unless of course the burst source resides within such a cluster.

Groups of galaxies may contain a significant fraction of the IGM as intragroup gas. \citet{s3} summarized the X-ray properties of groups of galaxies. The integrated radial column density based on an average over 23 groups gives ${\rm DM} \approx 150$ cm$^{-3}$ pc. From the mass function of groups and clusters of galaxies \citep{b1}, the present space density of such objects is $\approx 2.5\times 10^{-3}$ Mpc$^{-3}$, with the low mass groups numerically dominant. For a typical radius of $\approx 0.3$ Mpc, the characteristic line-of-sight length for encountering a group then is $\approx 1.4$ Gpc. On average, this yields an effective DM-distance ratio of $\approx 100$ cm$^{-3}$ pc Gpc$^{-1}$. Not surprisingly, this is comparable to that expected from a uniform, ionized IGM that contains essentially the full baryon content of the universe \citep{l2}, i.e. $\approx 280$ cm$^{-3}$ pc Gpc$^{-1}$. (Both values are valid for moderate redshifts.) The contribution from groups is highly uncertain as many groups do not show significant X-ray emission, either because they contain less gas, or because the gas within them is cooler, and therefore less luminous in the detected bands. The known hot gas in the Local Group probably contributes little to the observed dispersion measures. A hot corona of density $\approx 10^{-4}$ cm$^{-3}$ extending  perhaps 100 kpc from the Milky Way \citep{s2} would contribute only 10 pc cm$^{-3}$ to the DM.

The intragroup and intracluster gas have temperatures ranging from $\approx 0.8$ keV in groups to $\approx 8$ keV in rich clusters. Because the gas can be regarded as 
relativistically ``warm," it is worth examining whether this might be apparent in the resulting dispersion curves. For the low densities that are relevant, a simple expression for the group velocity can be obtained from the dispersion relation given by \citet{b3}:
\begin{equation} 
V_g = c\sqrt{1 - \Big(1- {{3kT}\over{2mc^2}}\Big)\Big({\omega_p\over\omega}\Big)^2}
\end{equation}  
For even the hottest rich clusters the relativistic correction is small and equivalent to decreasing the plasma frequency by 0.5\%. More importantly, the pulse delay retains the $\nu^{-2}$ behavior characteristic of cold dilute plasma. 

The directions toward bursts were examined for X-ray emission using SIMBAD, and in two cases nearby X-ray emission was found: The source [HFP 2000]704 \citep{h1} lies approximately 6 arc minutes east of the telescope beam coordinates corresponding to the maximum flux reported by \citet{l2} for FRB 010724. A direct relationship with this FRB seems unlikely, however, as the burst was also detected with beams west and north of the peak location, but not east. Also, the source 1XRS J232917.6-025430 \citep{v1} lies approximately 11 arc minutes from the telescope coordinates reported for FRB 110703 \citep{t1}. Clearly, these coincidences are not evidence for emission by the dispersing gas. Nevertheless, highly sensitive X-ray observations might be able to detect such emission if the dispersing gas is sufficiently hot and concentrated. 

\section{Conclusions}

Fast radio bursts exhibit quadratic dispersion curves that are consistent with the assumption of weak dispersion in a low density plasma in which the plasma frequency is well below the observation band, typically 1.2 -- 1.5 GHz. This rules out flares occurring near the photospheres of M stars or contact binaries in which most of the observed dispersion is produced in the overlying stellar coronae \citep{l1}. In general, observers are urged to fit dispersion curves of FRBs to a general expression, based on equation (6), in order to detect or limit strong dispersion caused by dense plasma, possibly associated with the source. 

Evidently then the observed dispersion occurs in a low density medium along the line of sight. The interstellar medium, however, can not account for the observed DMs, and thus we must conclude that most of the observed dispersions are extragalactic \citep{l2,t1}, at least in the cases of events observed at high galactic latitude. Current distance estimates are based on a uniform IGM that contains most of the baryonic mass of the universe. This medium is certainly concentrated in large scale structure, however, and much of it may be observable as X-ray emitting intracluster and intragroup gas. It is therefore important to identify line of sight structures in the directions of known FRBs. Especially valuable will be determination of redshifts of foreground structures, whether through optical observations of galaxies or X-ray observations of intervening gas, in order to begin to probe the distribution of the IGM.

\acknowledgments

I thank Dr. C. Bennett, J. Daugherty, and S. L. O'Dell for useful discussions. This work was supported in part by the Glaxo-Wellcome Endowment at UNC-Asheville.

\end{document}